\documentclass[twocolumn,preprintnumbers,showkeys]{revtex4}
\usepackage{xcolor}
\usepackage{graphicx}
\usepackage{graphicx,epstopdf}
\usepackage{dcolumn}
\usepackage{bm}
\usepackage{amsmath,mathrsfs}
\usepackage{amsthm,amssymb,amsmath,mathrsfs,mathtools}
\usepackage[mathscr]{euscript}
\usepackage[colorlinks=true,citecolor=blue]{hyperref}

\begin{document}

\title{Dynamical robustness of complex networks subject to long-range connectivity}

\author{Soumen Majhi}
\email{soumen.majhi91@gmail.com}
\affiliation{Department of Mathematics, Bar-Ilan University, Ramat-Gan, 5290002, Israel}

\begin{abstract}
In spite of a few attempts in understanding the dynamical robustness of complex networks, this extremely important subject of research is still in its dawn as compared to the other dynamical processes on networks. We hereby consider the concept of long-range interactions among the dynamical units of complex networks and demonstrate \textit{for the first time} that such a characteristic can have noteworthy impacts on the dynamical robustness of networked systems, regardless of the underlying network topology. We present a comprehensive analysis of this phenomenon on top of diverse network architectures. Such dynamical damages being able to substantially affect the network performance, determining mechanisms that boost the robustness of networks becomes a fundamental question. In this work, we put forward a prescription based upon self-feedback that can efficiently resurrect global rhythmicity of complex networks composed of active and inactive dynamical units, and thus can enhance the network robustness. We have been able to delineate the whole proposition analytically while dealing with all $d$-path adjacency matrices, having an excellent agreement with the numerical results. For the numerical computations, we examine scale-free networks, Watts-Strogatz small world model and also Erdös-Rényi random network, along with Landau-Stuart oscillators for casting the local dynamics.
\end{abstract}


\maketitle


\section{Introduction}

Complex networks provide a powerful tool to understand diverse processes taking place in a wide range of complex systems \cite{rev1,rev2}. A major achievement of the study of networks is that it is capable of describing collective scenarios having strong applications in large complex systems. The study of dynamical evolution of networked systems leads to a better explanation of a variety of complex processes, ranging from different aspects of synchronization \cite{newsyn1,newsyn2,newsyn3,newsyn4,newsyn5,newsyn6} and evolution of cooperation \cite{newegd1,newegd2,newegd3,newegd4,newegd5} to contagion processes \cite{newepi1,newepi2,newepi3,newepi4}. Among all these phenomena, network robustness has been in the forefront as it explores the issue of networks' ability to stand against perturbations. Networks' structural robustness deals with the structural perturbations by means of progressive removal (addition) of links or nodes, and looks into the presence (appearance) of giant connected component in the network \cite{top1,top2,top10,top5,top9}. Besides, dynamical robustness investigates the global dynamism of the whole ensemble subject to local perturbations in the dynamics of the nodes. In literature, this is realized by considering damaged networks comprising of active (oscillating) and inactive (non-oscillating) dynamical units, and then by inspecting how long the global oscillation persists in the network depending upon the increasing number of inactive units \cite{dyn1,dynpr}. 

\par The relevance of such studies in natural and man-made scenarios is obvious. Not only living systems experience deterioration due to aging or environmental influence or accidental events, but also several man-made systems are bound to undergo local failures. Whenever such degradation surpasses certain limit, the functioning of the whole ensemble ceases. This requires one to study the extent of resistance of the networked systems against these kinds of disruptions. As far as the neuronal systems are concerned, both individual neurons and neuronal ensembles exhibit strong tendency of taking part in rhythmic activity \cite{ap1,ap2}. Oscillatory activity in neurons plays crucial role in processing neural information and temporal coordination of processes underlying visual sensing, cognitive tasks, memory \cite{ap3,ap5}. Firing of neurons is also fundamental to central pattern generators. So, cessation of oscillation in neurons can have outright effect on the fundamental neural processes \cite{apn1}. On the other hand, our planet is facing extinction of species at faster rates than ever. This disaster predominantly owes to the climate change, geologic catastrophes, extreme use of natural resources, and assemblage complexity, affecting the surrounding ecosystems in a large scale \cite{ap6,ap9}. In ecological networks, extinction of patches in the metapopulation can thus lead to dramatic changes in its persistence \cite{ap8,ap10,ap11,ap11b}. Also, accurate functioning of cardiac to respiratory systems \cite{ap12} to physiological processes like cell necrosis within organs \cite{ap13} demands oscillatory dynamics. Power-grid networks compulsorily need stable synchronized rhythmic activity \cite{ap14,ap15}.

\par  The study of dynamical robustness being highly pertinent to several natural and real-world processes, in the last decade, efforts have been made to interpret how intricate connectivity patterns influence the dynamical robustness of the networked systems. For example, starting with the seminal work of Daido et al. \cite{dyn1}, the ref. \cite{dyn2} also dealt with globally coupled networked systems. From locally coupled \cite{dyn4} to scale-free \cite{dyn6} and multilayer networks \cite{dyn5} have been assumed to frame the interaction among the constituents. Even, mean-field coupled network \cite{bisw}, weighted network \cite{arn}, additional repulsion \cite{bid,bisw2} and time-delayed coupling \cite{dyn10} has been taken into account that affect aging of networks. 

\par  As the research on network theory having potentiality of describing natural scenarios, is developing so fast, it has become increasingly necessary to move beyond simple network structures and explore more sophisticated but realistic connectivity patterns. In this regard, the interaction structure of networked systems is found to form not only by the direct links among the network nodes, but also by the long-range interactions arising through other many existing paths connecting the nodes. Long-range interacting systems  \cite{lrr4,lrr5,lrr1,lrr2,lrr6} are omnipresent and have emerged to be one of the most prosperous areas of research in complex systems ranging from mechanical to biological networks \cite{lr2}. Being specific, long-range interaction owing to power-law decay has been examined for biological networks \cite{lr3}, hydrodynamic interaction of active particles \cite{lr5,lr6}, ferromagnetic spin models \cite{lr4}, Rydberg atoms \cite{lr8}, nuclear spins in solid-state systems \cite{lr7}, plasmas \cite{lr9} etc. Due to the extensive applicability of long-range interacting systems, it is necessary to understand the dynamical behavior of networks of oscillators subject to such kind of interactions, specifically the one related to the robustness of networks.
Nevertheless, to the best of our knowledge, there hasn't been any attempt to probe the theory of dynamical robustness in complex networks put through long-range connections. Through this article, we show \textit{for the first time} how the dynamical robustness of coupled systems evolving on complex networks gets affected based on the degree of the long-range connectivity that essentially depends upon the underlying network topology (responsible for the direct interactions among the nodes). \textit{Thus, we essentially aim at this unaccomplished area of research and study how robust a complex network subject to long-range interaction is, and how can the extent of this robustness be enhanced? }  

\par As far as the other collective dynamical behaviors in long-range interacting ensemble of dynamical systems are concerned, there exists investigation of the synchronization phenomenon in phase oscillators \cite{phase,scalefr}, discrete dynamical systems \cite{phase2} and also in chaotic systems \cite{chalri}. Network synchronizability owing to such interaction for both Mellin and Laplace transformed versions is studied in the ref. \cite{siam}. Chimera-like states along with oscillation suppression is also discussed recently \cite{tb1,sathiya}. Persistence of meta-population for long-range dispersal is also reported \cite{tb2}. Taken together, one must note here that not only the issue of network robustness is ignored, but also in most of the cases the underlying network framework has been presumed to be regular networks and not the complex ones.        

\par Here we must note that only a very few works exist in the literature which comes with ideas that reduces the vulnerability of the networks and enhances their robustness against such dynamical deterioration. The existing ones include the ref. \cite{dyn13} dealing with a diffusion controlling parameter for enhancing robustness. Addition of supporting oscillators to damaged networks can develop the survivability of the networks \cite{dyn7}. Random errors in the parameters of the local dynamical units can also bring global rhythmicity back \cite{dyn15}. Recently, the authors of \cite{we1} studied that a \textit{global} feedback from all the nodes helps in augmenting dynamical persistence. Mismatch in coupling scenario also affects the robustness of networks positively \cite{we2}. But, a feedback from all the nodes of the network irrespective of the network topology readily implies a kind of global interrelational scenario, and may not make much sense in many occasions. On the other hand, addition of new systems to the network increases the effective size of it. In the present work, we show that a simple \textit{self-feedback} is quite sufficient in significantly enhancing the dynamical robustness of any aging network even if it's subject to long-range connections among the nodes. We thus do so without disturbing the network size, the rate of diffusion among the constituents or the local parameters of the dynamical units. 
\par In the present article, in order to introduce our work, we first illustrate the concept of long-range interactions among the nodes in a network (cf. Sec II). Then we discuss our main findings on dynamical robustness, particularly the numerical results on diverse complex network architectures in Sec. IIIA, and the analytical approach towards finding the critical inactivation ratio for aging of the networks subject to long-range connectivity in Sec. IIIB. In Sec. IV, we introduce our mechanism for resurging dynamical activity and hence enhancing the robustness of such networks, elaborate the obtained numerical results, and validate them through analytical findings. Finally, we provide concluding remarks on our work in Sec. V. 


\section{Long-range interaction in networked systems}\label{sec2}

In large complex systems, interactions among its constituents can often be expressed in terms of a network representation through its adjacency matrix. To set the stage, we first define the concept of long-range interaction in networked systems, in terms of the adjacency matrices associated to different path lengths. Let $G=(V,E)$ be a network with $N$ number of nodes where $V$ and $E$ are the sets of nodes and links respectively, {\it i.e.} $V=\{1,2,\dots,N\}$, and $E\subset V\times V$ is the set of undirected links. Then the diameter of the network can be defined as $D=\max\big\{dist(j,k):j,k=1,2,\dots,N\big\}$. Here $dist(j,k)$ is the distance between the nodes $j$ and $k$, which is basically the length of any shortest path between $j$ and $k$.  

\par The $d$-path adjacency matrix ${A}^{[d]}$ of the network is then described as the following symmetric $N\times N$ matrix,
\[
{A}_{jk}^{[d]}= 
\begin{cases}
	1 & \text{if}~dist(j,k)=d, \\
	0  & \text{otherwise}.
\end{cases}
\]

\begin{figure}[ht]
	\centerline{\includegraphics[scale=0.300]{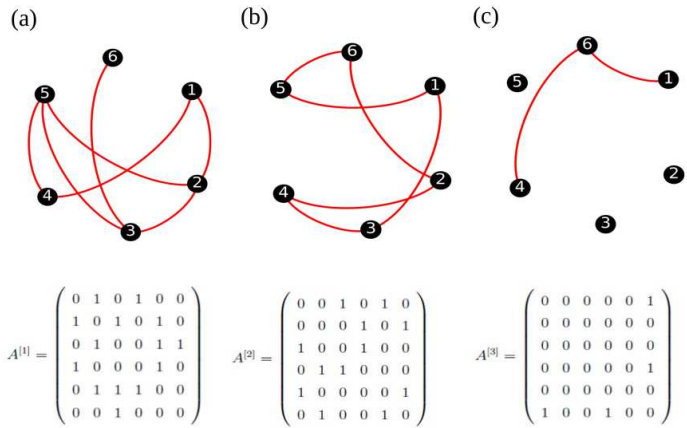}}
	\caption{An exemplary network with long-range connections: The $d$-path networks and the corresponding adjacency matrices are respectively depicted in the upper and lower panels: (a) $d=1$, {\it i.e.}, the original network, (b) $d=2$, {\it i.e.}, the $2$-path network extracted from (a), and (c) $d=3$, {\it i.e.}, the $3$-path network extracted from (a).}
	\label{fig0}
\end{figure}

\par For a better explanation of all the $d$-path networks and hence the long-range interactions among the nodes, we consider a small exemplary network of $N=6$ nodes with $D=3$, and plot the corresponding $d$-path $~(d=1,2,3)$ networks together with the corresponding adjacency matrices in Fig. \ref{fig0}. Figure \ref{fig0}(a) portrays the $1$-path network ({\it i.e.} the fundamental network structure associated to the direct interactions), along with the adjacency matrix ${A}^{[1]}$. As can be seen, here $(1,2)\in E$ and $(2,3)\in E$ but $(1,3)\notin E$, thus $dist(1,3)=2$ and hence ${A}_{13}^{[2]}=1$. Proceeding this way, the $2$-path network can be constructed (cf. Fig. \ref{fig0}(b)). Noting that $(3,6)\in E$, there is a path of length three from the node $1$ to the node $6$, namely $(1,2)\rightarrow(2,3)\rightarrow(3,6)$, but there exists no path from $1$ to $6$ having length less than three. So, we have $dist(1,6)=3$, yielding ${A}_{16}^{[3]}=1$. Also, there exists another path of length three in the network, which is from the node $4$ to $6$. Hence, the $3$-path network consists of only two links $(1,6)$ and $(4,6)$. The remaining three nodes $2,3$ and $5$ are isolated in the $3$-path network. The pictorial view of this network is presented in Fig. \ref{fig0}(c), with the corresponding adjacency matrix ${A}^{[3]}$. This description of long-range interactions readily suggests that the concept of a decaying interaction strength in terms of the distance between the considered nodes, is quite relevant and realistic. This is more so if power-law decay is assumed to model the scenario, as ascertained by the preceding references in the Introduction section.

\par Let us now assume that each node in the network is affiliated with a $l$-dimensional dynamical system. Then the dynamical evolution of the $j$-th node in the network owing to long-range interaction can be written as follows, 
\begin{equation}
	\begin{array}{lcl}\label{eqq1}
		\dot{\bf x}_j={\bf f}({\bf x}_j)+\dfrac{1}{N}\sum\limits_{d=1}^{D}\sigma_d\sum\limits_{k=1}^{N}{A}^{[d]}_{jk}\Gamma({\bf x}_k-{\bf x}_j),\\
		~~~~~~~~~~~~~~~~~~~~~~~~~~~~~~~~~~~j= 1,2,\cdots,N,
	\end{array}
\end{equation}
where ${\bf x}_j$ corresponds to the $l$-dimensional state variable of the $j$-th node, ${\bf f}:{\mathbb{R}}^{l}\rightarrow {\mathbb{R}}^{l}$ is the vector field representing the dynamics of each node in absence of any interaction, $\sigma_d$ is the coupling strength between the $j$-th and $k$-th nodes if $dist(j,k)=d$. In other words, the $j$-th and $k$-th nodes interact with strength $\sigma_d$ if they are $d$-path connected, where $d=1,2,\dots,D$, $D$ being the diameter of the network. Essentially, this means that the strength of interaction between certain nodes strictly depends on the distance between them. For instance, the coupling strength associated with each link for the $1$-path network in Fig. \ref{fig0}(a) is $\sigma_1$ whereas those associated to each link of the networks in Fig. \ref{fig0}(b) and Fig. \ref{fig0}(c) are $\sigma_2$ and $\sigma_3$ respectively. ${A}^{[d]}_{jk}$ are the entries of the $d$-path adjacency matrix ${A}^{[d]}$ and $\Gamma$ is the inner coupling matrix defining which state variables of the nodes cause the interaction among them.

\par In the next, we carry out our inspection while considering the local dynamical units as the Stuart-Landau (SL) oscillators \cite{dyn1} as follows, \\
\begin{equation}
	\begin{array}{lcl} \label{eqq2}
		{\bf f}(z_j) =(\alpha_j + i\Omega - |z_j|^2)z_j;
		~~j= 1,2,\cdots,N,
	\end{array}
\end{equation}
where $\alpha_j$ is an internal parameter of the $j$-th system defining the distance from a Hopf bifurcation, $\Omega$ is the natural frequency of oscillation and $i=\sqrt{-1}$. While isolated, $j$-th unit possesses a stable limit cycle $\sqrt{\alpha_j} e^{i \Omega t}$ whenever $\alpha_j>0$, and remains stable in the trivial fixed point regime $z_j=0$ if $\alpha_{j}\le 0$. 
Therefore, we contemplate with certain values $\alpha_j=a>0$ and $\alpha_j=-b<0$ respectively in order to specify the active and inactive oscillators in the network.

\par We then ensue the process \cite{dyn1} that assumes the inactivation ratio (ratio of non self-oscillatory units) $p$ which is the ratio of the number of inactive nodes and the total number of nodes in the network. While increasing, whenever $p$ exceeds a certain threshold $p_c$ (say) (generally depending on all the network parameters), the global oscillation of the network dies out. 
We explain the scenario of dynamical robustness of the network by looking at the the aging transition in terms of the normalized order parameter $Z = \frac{|\bar{Z}(p)|}{|\bar{Z}(0)|}$, where $\bar{Z}=\dfrac{1}{N}\sum\limits_{j=1}^{N}z_j$ specifies the intensity of global oscillation in the networked system and $Z$ is the normalized value of it.

\section{Results}

\subsection{Numerical Outcomes}

\begin{figure*}[ht]
	\centerline{
		\includegraphics[scale=0.700]{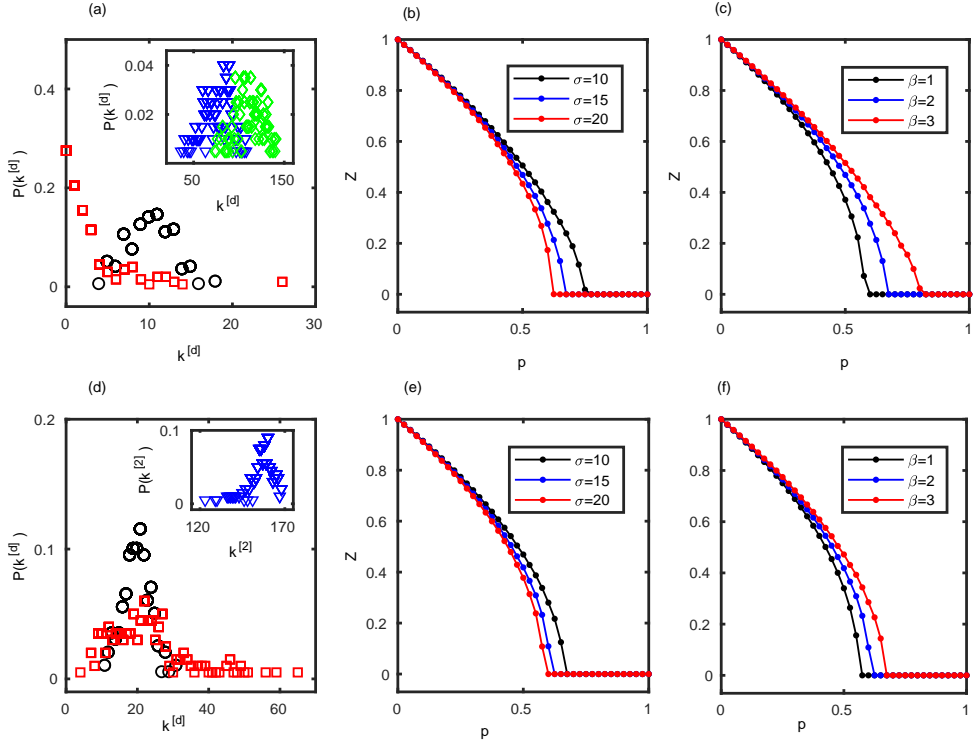}}
	\caption{ The degree distribution $P(k^{[d]})$ corresponding to each $d$-path network for (a) $q=0.05$ with $d=1,2,3,4$ (the plots associated to $d=1,4$ are in inset), and (d) $q=0.10$ with $d=1,2,3$ (the plot associated to $d=2$ is in inset). The order parameter $Z$ as a function of the inactivation ratio $p$ for (b,e) different coupling strength $\sigma$ with $\beta=2$ fixed, and (c,f) different decay rate $\beta$ with $\sigma=15$ fixed. The upper panel (b,c) corresponds to the $G(N,q)$ model with $q=0.05$ whereas the lower panel (e,f) corresponds to the $G(N,q)$ model with $q=0.10$.}
	\label{fig1}
\end{figure*}

\par As hinted above, in this paper, we mainly concentrate on the role of power-law decay of the interaction strength with respect to the distance between the concerned nodes. Thus, if $\sigma_d$ be the coupling strength between the nodes having distance $d~(d=1,2,\dots,D)$, then $\sigma_d=\dfrac{\sigma}{d^{\beta}}$, $\beta$ being the power-law exponent regulating the decay rate. We start our investigation considering the underlying network endowed with ER random architecture, {\it i.e.}, we choose the $G(N,q)$ graph model \cite{er2} with $N=200$ as the number of nodes and $q$ as the connection probability. Without loss of generality, $a=1$, $b=1$ and $\Omega=3$ are considered throughout the work, where the fifth order Runge-Kutta Fehlberg scheme has been used to integrate the networked dynamical system.

\par We note that, in the present work, the characteristic of long-range connectivity is the fundamental mode of interaction among the nodes. So, in order to have a perception of what happens with the network structure in the basic level, we first choose $q=0.05$ and look at the degree distribution $P(k^{[d]})$ corresponding to each $d$-path network where $d=1,2,3,4$, the diameter $D$ of the underlying network being $D=4$. Figure \ref{fig1}(a) depicts $P(k^{[d]})$ as functions of $k^{[d]}$, from which one can perceive that not only the $1$-path network associated to the direct communications follow Poisson distribution, but also the other $d(>1)$-path networks can possess similar distributions. Next, we keep the value of $\beta$ fixed at $\beta=2$, and plot $Z$ in respect of the increasing inactivation ratio $p$ $(0\le p\le 1)$ for different values of $\sigma$ (cf. Fig. \ref{fig1}(b)). As can be observed, the order parameter $Z$ decreases to zero at $p_c\sim 0.75$ whenever $\sigma=10$, so that the networked system gets stabilized to the trivial fixed point for $p\ge p_c$ and hence the aging transition takes place. For a larger $\sigma=15$, the transition occurs at a smaller $p_c\sim 0.67$, as in this case the nodes interact with higher strength. For a further increment in $\sigma$ to $\sigma=20$, aging happens much before than the previous two cases, particularly at $p_c\sim 0.62$. This basically indicates that if the coupling strength is that strong, even if only $62\%$ of nodes become inactive the global oscillation of the network can die out. In our work we focus mainly on the role of power-law decay of the interaction strength with respect to the distance between the concerned nodes. So, we now elaborately discuss the role of this power-law decay. We fix the value of $\sigma=15$, and check what role the power-law decay parameter $\beta$ plays. In Fig. \ref{fig1}(c), we plot $Z$ against $p$ for $\beta \in \{1,2,3\}$. When $\beta=1$, starting from unity (for $p=0$), $Z$ drops down to zero at $p_c\sim 0.60$. However, for higher values of $\beta$, the critical inactivation ratio $p_c$ also increases, in contrast to the scenario of the coupling strength $\sigma$. For instance, $p_c\sim 0.67$ for $\beta=2$, and $p_c\sim 0.81$ if $\beta=3$. Therefore, the appearance of aging delays and hence the network becomes more robust to the units' inactivation for the increasing power-law decay rate.      
\par We next perform a similar analysis on the network but with a different connection probability $q=0.10$. This time, $q$ being higher, the diameter $D$ reduces to $D=3$. The degree distributions $P(k^{[d]})~(d=1,2,3)$ are presented in Fig. \ref{fig1}(d). With $\beta=2$ being fixed as before, the phenomenon of aging transition for $\sigma \in \{10,15,20\}$ is discernible from Fig. \ref{fig1}(e). Here the global oscillation of the entire network vanishes at $p_c\sim 0.67,0.62,0.60$ (corresponding to these $\sigma$ values), all of which are smaller than their respective counterparts in case of $q=0.05$ (cf. Fig. \ref{fig1}(b)), owing to the fall of diameter of the network. On the other hand, the extent of dynamical robustness can be observed for varying $\beta \in \{1,2,3\}$ in  Fig. \ref{fig1}(f), with $\sigma=15$. Here again, the realized critical $p_c$ values are smaller than the values obtained earlier for $q=0.05$. But, increasing values of $\beta$ help the network to sustain global dynamism as before.  

\begin{figure}[ht]
	\centerline{\includegraphics[scale=0.5900]{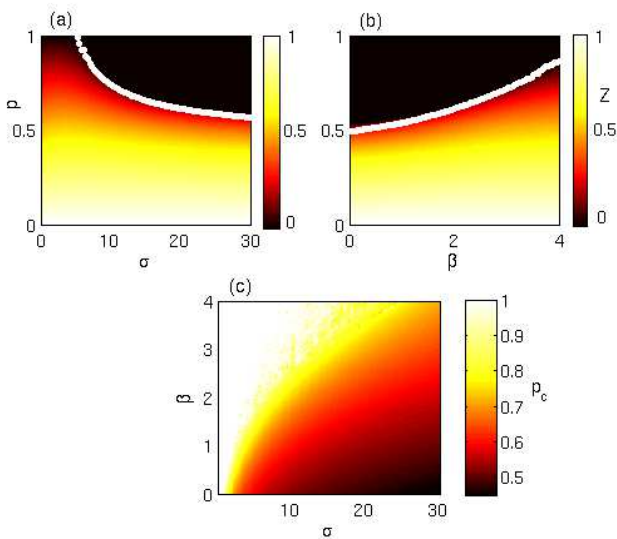}}
	\caption{The order parameter $Z$ as a function of : (a) the inactivation ratio $p$ and the coupling strength $\sigma$ for which $\beta=2$ is fixed, (b) the inactivation ratio $p$ and the decay rate $\beta$ for which $\sigma=20$ is fixed; (c) The critical $p_c$ with respect to the simultaneous variation of $\sigma$ and $\beta$. The white curves in (a) and (b) are the analytically obtained $p_c$ values. }
	\label{fig2}
\end{figure}

\par We now illustrate the above scenario in more details. To begin with, in Fig. \ref{fig2}(a), we settle for $q=0.05$ and depict the numerically found order parameter $Z$ for simultaneous variation in the inactivation ratio $p$ and the coupling strength $\sigma \in [0,30]$, where the long-range exponent $\beta$ has been kept fixed at $\beta=2$. For this range of $\sigma$, one can see the transition of $Z$ from non-zero values to zero (in case aging transition occurs) as $p$ increases. The higher values of $\sigma$ always lead to a faster dynamical collapse of the network. Upon this phase diagram, we then plot the analytically obtained (discussed later) critical inactivation ratio $p_c$ for these $\sigma \in [0,30]$. This curve of $p_c$ values has considerable agreement with the numerical outcome, and thus appear as a sufficiently precise boundary curve separating the aging and non-aging scenarios. Figure \ref{fig2}(b) presents the alteration in the values of $Z$ as a result of increasing $p$ and $\beta \in [0,4]$, where $\sigma=20$ is fixed. It is visible that with increasing values of the exponent $\beta$ the values of the critical inactivation ratio $p_c$ also increases strictly. On the contrary to the impact of $\sigma$ on the robustness, here increasing $\beta$ hinders aging, which is thus clear from Fig. \ref{fig2}(b). We plot the analytical $p_c$ values over this figure, that match with the numerical $p_c$ values quite well. For a better perception, next we plot $p_c$ in the $\sigma-\beta$ parameter plane, as in Fig. \ref{fig2}(c). This figure explains the synergy between these two parameters characterizing how robust the network is against local units' dynamical degradation. As can be observed from the figure, the critical inactivation ratio $p_c$ remains $1$ for small interaction strength $\sigma$ no matter what the value of $\beta$ is. This is simply because of the insufficiency of the coupling strength that restrains the network from any aging. However, as $\sigma$ increases the network starts experiencing aging transition and the $p_c$ value decreases. On the other hand, $p_c$ increases for higher $\beta$ as higher $\beta$ weakens the strength of the long-range communications. Thus the networked system is most vulnerable to this dynamic degradation when the power-law decay rate $\beta$ is low and the interaction strength $\sigma$ is high.       
\par
\begin{figure}[ht]
	\centerline{\includegraphics[scale=0.4500]{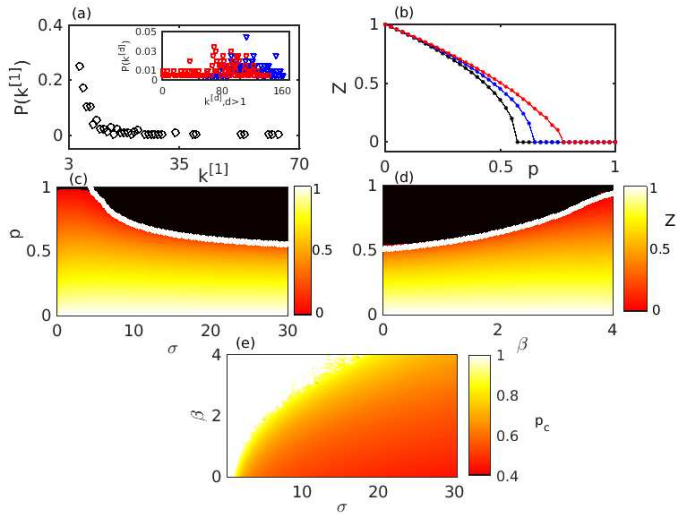}}
	\caption{(a) The degree distribution $P(k^{[d]})$ corresponding to $d$-path network for $d=1,2,3$ (the plots associated to $d=2,3$ are in inset);(b) The order parameter $Z$ with respect to the inactivation ratio $p$ for different values of the decay rate $\beta=1,2,3$ with $\sigma=15$ fixed. The order parameter $Z$ as a function of : (c) the inactivation ratio $p$ and the coupling strength $\sigma$ for which $\beta=2$ is fixed, (d) the inactivation ratio $p$ and the decay rate $\beta$ for which $\sigma=15$ is fixed; (e) The critical inactivation ratio $p_c$ in the $\sigma-\beta$ parameter plane. The white curves in (c) and (d) are the analytically obtained $p_c$ values.}
	\label{fig3}
\end{figure}

 Now will be discussing the scenario of long-range communication over a complex network with significant heterogeneity in its connectivity pattern, and the dynamical robustness of such networks thereafter. Regarding the degree distribution, a scale-free topology possesses a power-law $P(k)\sim k^{-\gamma}$ distribution, $P(k)$ being the probability of finding a node with degree $k$ and $\gamma$ the power-law exponent. Following the Barab\'{a}si-Albert mechanism \cite{sfn} of generating scale-free models, we construct such one with $N=200$ and $\gamma=3$. Figure \ref{fig3}(a) describes the degree distribution $P(k^{[1]})$ of the underlying network associated to the direct $1$-path connections. The other degree distributions $P(k^{[d]})$ corresponding to the indirect $d~(\ge 2)$-distant pathways are also shown in the inset of the same figure. As can be seen, the $1$-path network follows power-law degree distribution, but we don't see such heterogeneity in the distributions $P(k^{[d]});~d=2,3$. 

\par Considering $\sigma=15$, Fig. \ref{fig3}(b) marks how $Z$ changes against increasing inactive elements (in terms of $p$) for different intensities of long-range interactions (in terms of $\beta$). With $\beta=1$, $Z$ drops down to zero and hence the network loses its robustness entirely at $p_c \sim 0.57$. For higher $\beta=2$, this transition occurs later, precisely at $p_c \sim 0.65$. Even higher $\beta=3$ results in a more delayed aging and the networked system collapses for $p\ge p_c \sim 0.77$. Figure \ref{fig3}(c) comes with $Z$ when $0 \le p \le 1$ and $0\le \sigma \le 30$ both change together for $\beta=2$, whereas Fig. \ref{fig3}(d) uncovers the extent of robustness in the $\beta-p$ parameter plane for $\beta \in [0,4]$ and $\sigma=15$. The former one shows the trend of decreasing robustness, and conversely the latter one depicts the tendency of increased robustness. Both the scenarios are verified with the analytically found $p_c$ values plotted on the respective phase diagrams. These $p_c$ curves essentially distinguish the regimes of aging and no-aging, which are in well agreement with the numerical results. Finally, we reveal how $p_c$ alters in the $\sigma-\beta$ parameter space in case of scale-free networks with all the parameters being same as before (cf. Fig. \ref{fig3}(e)). From the figure, the simultaneity of $\sigma$ and $\beta$ on the phenomenon of robustness can be understood. The critical inactivation ratio $p_c$ varies, precisely between $0.4$ and $1.0$, as $\sigma$ and $\beta$ change. Increasing $\sigma$ lowers the $p_c$ value whereas increasing $\beta$ lessens the possibility of aging and hence $p_c$ increases. The phase diagram explains this trade-off quite well.         

\par Let us now get back to one more much celebrated network model with homogeneous degree distribution. We, particularly, move on to analyze this phenomenon briefly for small-world networks. The seminal work of Watts et al. \cite{swn} came up with the idea of small-worldness in the connectivity pattern of networks ranging from social to neuronal networks, reckoning high clustering and low diameter. In their model, such small-worldness emerge as a result of simple random rewiring of links (with probability $p_{r}$, say) from a lattice model, interpolating between the two limiting cases of regular ($p_{r}=0$) and random ($p_{r}=1$) networks. 
\begin{figure}
	\centerline{\includegraphics[scale=0.4300]{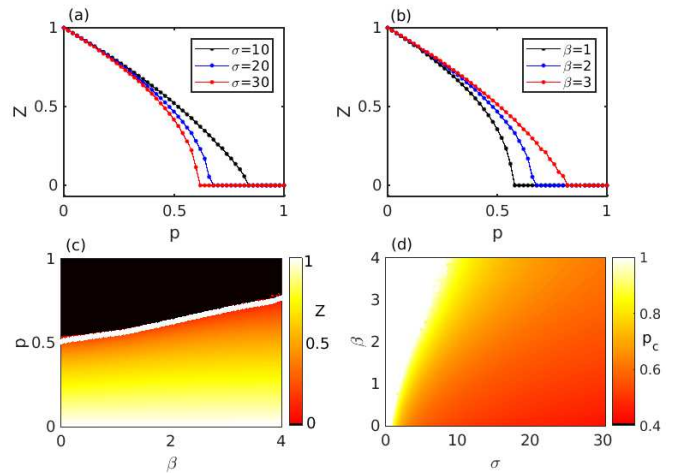}}
	\caption{The order parameter $Z$ as a function of the inactivation ratio $p$ for (a) different coupling strength $\sigma$ with $\beta=2$ fixed and (b) different decay rate $\beta$ with $\sigma=20$ fixed, for a small-world network model with the average degree $\langle k^{[1]} \rangle=10$ and the rewiring probability $p_{r}=0.1$. For a different small-world model possessing $\langle k^{[1]} \rangle=20$, $p_{r}=0.1$ : (c) variation of $Z$ with respect to $p$ and $\beta$ for which $\sigma=15$ is fixed; (d) The numerically obtained critical $p_c$ with respect to the simultaneous variation of $\sigma$ and $\beta$. The white curve in (c) corresponds to the analytically obtained $p_c$ values. }
	\label{fig4}
\end{figure}
\par  Figure \ref{fig4} (a) shows how $Z$ changes with respect to $p$ for three different values of $\sigma$ while $\beta=2$, for a small-world network with the average degree $\langle k^{[1]} \rangle=10$ and the rewiring probability $p_{r}=0.1$. Whenever, $\sigma=10$, $Z$ becomes zero for all $p \ge p_c\sim0.84$. But, if $\sigma=20$ and $30$, aging occurs at $p_c \sim0.68$ and $0.62$, implying earlier aging and hence less robustness of the ensemble. Similarly, Fig. \ref{fig4} (b) explains aging in case of varying $\beta$. Increasing $\beta=1,2,3$ eventually leads to $p_c \sim 0.58,0.68,0.82$ respectively. We also go for depicting $Z$ for varying $\beta$ and $p$ simultaneously, with $\sigma=15$ fixed and $\langle k^{[1]} \rangle=20$, $p_{r}=0.1$. This reveals to what extent $\beta$ matters, and quantifies the drop down of $p_c$ values for increasing $\beta \in [0,4]$ (cf. Fig. \ref{fig4} (c)). The outcome is then well validated through the analytical result (discussed below) of $p_c$. Lastly, we present the alteration in the $p_c$ values when both $\sigma \in [0,30]$ and $\beta \in [0,4]$ are varied. The opposing impact of these two parameters on aging is conspicuous from this diagram.    

\subsection{Analytical Treatment}

\par As mentioned earlier, one of the most important acquirements of network theory in the last decades is the demonstration of significant degree heterogeneity in complex systems. In order to generally deal with such heterogeneous networks (the approach considers the homogeneous cases as special cases and hence works perfectly), we follow the degree-weighted mean field approximation \cite{dwmf}. Consequently, the original system (Eq. (\ref{eqq1}) for Eq. (\ref{eqq2})) can be approximated as

\begin{equation}
	\begin{array}{lcl} \label{eqq3}
		\dot{z_j} = (\alpha + i\Omega - |z_j|^2)z_j + \dfrac{1}{N}\sum\limits_{d=1}^{D}\sigma_d{k_j}^{[d]}\Big[(1-p)M_A(t)\\~~~~~~~~~~~~~~~~~~~~~~~~~~~~~~~~~~~~~~+pM_I(t)-z_j\Big].\\
	\end{array}
\end{equation}

Here,
\begin{equation}
	\begin{array}{lcl} \label{eqq4}
		M_A(t) = \dfrac{\sum\limits_{d=1}^{D}\sigma_d\sum\limits_{j\in S_A} {k_j}^{[d]}z_j(t)}{\sum\limits_{d=1}^{D}\sigma_d\sum\limits_{j\in S_A} {k_j}^{[d]}}
	\end{array}
\end{equation}
is the degree-weighted mean field corresponding to the active set of dynamical units. Similarly,

\begin{equation}
	\begin{array}{lcl} \label{eqq5}
		M_I(t) = \dfrac{\sum\limits_{d=1}^{D}\sigma_d\sum\limits_{j\in S_I} {k_j}^{[d]}z_j(t)}{\sum\limits_{d=1}^{D}\sigma_d\sum\limits_{j\in S_I} {k_j}^{[d]}}
	\end{array}
\end{equation}  
is the degree-weighted mean field for the inactive group of systems. Also ${k_j}^{[d]}(j=1,2,...,N)$ is the degree of the $j$-th node associated to the $d$-path network. Let us now assume that the state variables can be written as $z_j(t)=r_j(t)e^{i(\Omega t+\theta)}$, $r_j$ being the amplitude and $\theta$ being the phase shift. On replacing this in Eq. \eqref{eqq3}, we have
\begin{equation}
	\small{\begin{array}{lcl}  \label{eqq6}
			\dot{r_j} = \Big(\alpha_j - \dfrac{1}{N}\sum\limits_{d=1}^{D}\sigma_d{k_j}^{[d]}-r_j^2\Big)r_j + \dfrac{1}{N}\sum\limits_{d=1}^{D}\sigma_d{k_j}^{[d]}\Big[(1-p)R_A(t)\\~~~~~~~~~~~~~~~~~~~~~~~~~~~~~~~~~~~~~~~~~~+pR_I(t)\Big],\\
	\end{array}}
\end{equation}
where
\begin{equation}
	\begin{array}{lcl} \label{eqq7}
		R_A(t) = \dfrac{\sum\limits_{d=1}^{D}\sigma_d\sum\limits_{j\in S_A} {k_j}^{[d]}r_j(t)}{\sum\limits_{d=1}^{D}\sigma_d\sum\limits_{j\in S_A} {k_j}^{[d]}},\\\\
		R_I(t) = \dfrac{\sum\limits_{d=1}^{D}\sigma_d\sum\limits_{j\in S_I} {k_j}^{[d]}r_j(t)}{\sum\limits_{d=1}^{D}\sigma_d\sum\limits_{j\in S_I} {k_j}^{[d]}}.
	\end{array}
\end{equation}
We further assume that in the stationary oscillatory regime, $R_A(t)$ and $R_I(t)$ are time-independent. Then the phase transition from the oscillatory ($R_A,R_I>0$) regime to the non-oscillatory ($R_A=R_I=0$) regime occurs owing to the change in the stability of the fixed point at the origin. The stability is governed by the following Jacobian matrix 
$$J_0=\begin{pmatrix}
	\dfrac{\partial G_A(R_A,R_I)}{\partial R_A} & \dfrac{\partial G_A(R_A,R_I)}{\partial R_I} \\
	\dfrac{\partial G_I(R_A,R_I)}{\partial R_A} & \dfrac{\partial G_I(R_A,R_I)}{\partial R_I}
\end{pmatrix}\Biggr\rvert_{R_A=R_I=0},$$
in which
\begin{equation}
	\begin{array}{lcl}  \label{eqq8}
		G_A(R_A,R_I)=\dfrac{\sum\limits_{d=1}^{D}\sigma_d\sum\limits_{j\in S_A} {k_j}^{[d]}r_j^*(R_A,R_I)}{\sum\limits_{d=1}^{D}\sigma_d\sum\limits_{j\in S_A} {k_j}^{[d]}},\\\\
		G_I(R_A,R_I)=\dfrac{\sum\limits_{d=1}^{D}\sigma_d\sum\limits_{j\in S_I} {k_j}^{[d]}r_j^*(R_A,R_I)}{\sum\limits_{d=1}^{D}\sigma_d\sum\limits_{j\in S_I} {k_j}^{[d]}}.
	\end{array}
\end{equation} 
Here the stationary amplitude $r_j^*$ is given by a positive real solution of the following equation:
\begin{equation}
	\begin{array}{lcl} \label{eqq9}
		r_j^3 - \Big(\alpha_j-\dfrac{1}{N}\sum\limits_{d=1}^{D}\sigma_d{k_j}^{[d]}\Big)r_j - \dfrac{1}{N}\sum\limits_{d=1}^{D}\sigma_d{k_j}^{[d]}\Big[((1-p)R_A\\
		~~~~~~~~~~~~~~~~~~~~~~~~~~~~~~~~~~~~~~~~~~~~+pR_I)\Big]=0.
	\end{array}
\end{equation}
Equation \eqref{eqq9} has only one positive real root if 
\begin{equation}
	\begin{array}{lcl} \label{eqq10}
		\alpha_j - \dfrac{1}{N}\sum\limits_{d=1}^{D}\sigma_d{k_j}^{[d]} <0,~\forall  j \in S_A. 
	\end{array}
\end{equation}

Now, in order to compute the entries of the Jacobian $J_0$, we differentiate Eqs. \eqref{eqq8} and \eqref{eqq9} with respect to $R_A$ and we obtain the $(1,1)$-th entry of $J_0$ as,
\begin{equation}
	\begin{array}{lcl} \label{eqq11}
		\dfrac{\partial G_A}{\partial R_A}\Bigr\rvert_{R_A=R_I=0} = \dfrac{\dfrac{(1-p)}{N}\sum\limits_{d=1}^{D}\sigma_d\sum\limits_{j \in S_A}{k_j}^{[d]}}{\sum\limits_{d=1}^{D}\sigma_d\sum\limits_{j \in S_A}{k_j}^{[d]}} \Big[\dfrac{L_j}{(L_j/N)-\alpha_j}\Big] \\ \\
		~~~~~~~~~~~~~~~~~~~=\dfrac{(1-p)\sum\limits_{d=1}^{D}\sigma_d\sum\limits_{j \in S_A}\Big[\dfrac{{k_j}^{[d]}L_j}{L_j-N\alpha_j}\Big]}{\sum\limits_{d=1}^{D}\sigma_d\sum\limits_{j \in S_A}{k_j}^{[d]}},
		
		\\ 
	\end{array}
\end{equation}
where $L_j=\sum\limits_{d=1}^{D}\sigma_d{k_j}^{[d]};~j=1,2,...,N$.\\
Now, denoting  $s^{[d]}=\langle k^{[d]} \rangle/(N-1)$ to be the link density for the $d$-path network ($d=1,2,...,D$), the following approximations hold in the large $N$ limit,

\begin{equation}
	\begin{array}{lcl} \label{eqq12}
		\sum\limits_{j \in S_A}{k_j}^{[d]} \simeq (1-p)s^{[d]}N^2,\\\mbox{and}~~~~\sum\limits_{j \in S_I}{k_j}^{[d]} \simeq ps^{[d]}N^2.\\
	\end{array}
\end{equation}

Thus from the Eqs. (\ref{eqq11}) and (\ref{eqq12}), we further have 

\begin{equation}
	\begin{array}{lcl} \label{eqq13}
		\dfrac{\partial G_A}{\partial R_A}\Bigr\rvert_{R_A=R_I=0}\simeq
		\dfrac{\sum\limits_{d=1}^{D}\sigma_d\sum\limits_{j \in S_A}\Big[\dfrac{{k_j}^{[d]}L_j}{L_j-N\alpha_j}\Big]}{N^2\sum\limits_{d=1}^{D}\sigma_d s^{[d]}}. 
	\end{array}
\end{equation}
If we now define
\begin{equation}
	\begin{array}{lcl} \label{eqq14}
		H(\sigma,\alpha)=\dfrac{\sum\limits_{d=1}^{D}\sigma_d\sum\limits_{j=1}^{N}\Big[\dfrac{{k_j}^{[d]}L_j}{L_j-N\alpha_j}\Big]}{N^2\sum\limits_{d=1}^{D}\sigma_d s^{[d]}},
	\end{array}
\end{equation}
then we can write
\begin{equation}
	\begin{array}{lcl} \label{eqq15}
		\dfrac{\partial G_A}{\partial R_A}\Bigr\rvert_{R_A=R_I=0}\simeq
		(1-p)H(\sigma,a). 
	\end{array}
\end{equation}
Proceeding in the same way, we will have
\begin{equation}
	\begin{array}{lcl} \label{eqq16}
		\dfrac{\partial G_A}{\partial R_I}\Bigr\rvert_{R_A=R_I=0}\simeq
		pH(\sigma,a),\\
		\dfrac{\partial G_I}{\partial R_A}\Bigr\rvert_{R_A=R_I=0}\simeq
		(1-p)H(\sigma,-b),\\
		\dfrac{\partial G_I}{\partial R_I}\Bigr\rvert_{R_A=R_I=0}\simeq
		pH(\sigma,-b).\\
	\end{array}
\end{equation}
Thus we arrive at 
$$J_0=\begin{pmatrix}
	(1-p)H(\sigma,a) & pH(\sigma,a)\\
	~~(1-p)H(\sigma,-b) &~~ pH(\sigma,-b)
\end{pmatrix}.$$

Near the phase transition the fixed point at $R_A=R_I=0$ changes its stability, which leads to the following critical inactivation ratio
\begin{equation}
	\begin{array}{lcl} \label{eqq17}
		p_c = \dfrac{H(\sigma,a)-1}{H(\sigma,a)-H(\sigma,-b)},
	\end{array}
\end{equation}
in which the definition of $H$ follows from Eq. (\ref{eqq14}). 

\section{Enhancing the dynamical robustness}
As proposed above, we here focus on the crucial issue of enhancing the dynamical robustness of networks,  specifically of blended network ensembles of active and inactive dynamical systems. In this section, we proceed to prescribe a simple yet highly efficient mechanism based upon the concept of self-feedback, that can effectively revive global rhythmicity in the network even if it is exposed to a large fraction of inactive units. Our objective will be to augment the robustness of the network and hence to keep global dynamism in the network as long as possible, by means of this approach.   

\par Feedback is the process through which output of a system is used back into the system as input, and is well-regarded as one of the most applicable concept in a number of scientific disciplines from physics, mathematics to biology and engineering \cite{fdb5}. In dynamical systems \cite{fbpla}, climate science, ecosystems, control theory \cite{fdb7} and even in the study of game theory \cite{gamefb1,gamefb2}, physiology \cite{fdbcan}, pattern formations, neuronal networks \cite{fdbplos}, the feedback theory has noteworthy contributions. Specifically, positive feedback has the ability of destabilizing dynamical systems from equilibrium, an impact which we intend to utilize here.


\begin{figure}
	\centerline{\includegraphics[scale=0.4600]{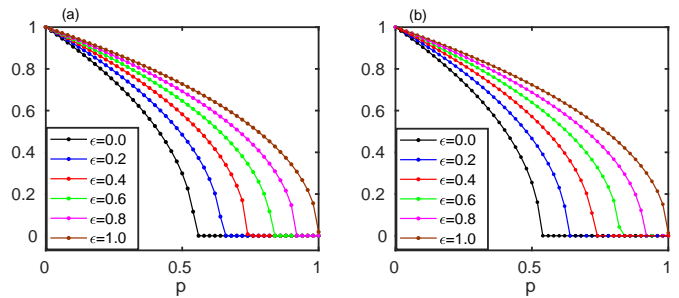}}
	\caption{The order parameter $Z$ with respect to the inactivation ratio $p$ for different self-feedback strength $\epsilon$ in case of small-world network models with (a) the average degree $\langle k^{[1]} \rangle=10$ and the rewiring probability $p_{r}=0.1$; (b) the average degree $\langle k^{[1]} \rangle=30$ and the rewiring probability $p_{r}=0.04$. In both the cases, $\sigma=30$ and $\beta=1$ are kept fixed.}
	\label{fig5}
\end{figure}

\par We start with the small-world network model with the purpose of scrutinizing the effect of the feedback strength $\epsilon$ on its robustness, for fixed values of the diffusive coupling parameter $\sigma$ and the long-range exponent $\beta$. For this, we deal with two different parameter set-ups in Fig. \ref{fig5}. For instance, we first work with a network where the average degree $\langle k^{[1]} \rangle=10$ and the rewiring probability $p_{r}=0.1$. We choose $\sigma=30$ and $\beta=1$, and see the difference in the network dynamics while changing $\epsilon$. For no feedback, $p_c$ is found to be $p_c\sim 0.56$ when $\epsilon=0$ . But, a small increment in $\epsilon$ to $\epsilon=0.2$ resurrects the global dynamical activity yielding $p_c\sim 0.66$, and hence the network robustness enhances (cf. Fig. \ref{fig5}(a)). If we increase the value of $\epsilon$ to $\epsilon=0.4$, aging turns out to be at $p_c \sim 0.75$. Further raise in the value of $\epsilon$ to $\epsilon=0.6,0.8,1.0$ deliver higher $p_c \sim 0.84,0.92,1.0$ respectively. In order to make sure that the results are qualitatively independent of the network model parameters, we next consider a small-world model in which $\langle k^{[1]} \rangle=30$ and $p_{r}=0.04$. In this case, for $\epsilon=0$, the $p_c$ appears to be $p_c \sim 0.54$, as in the order parameter (cf. Fig. \ref{fig5}(b)). An introduction of self-feedback with strength $\epsilon=0.2$ helps significantly in reviving global oscillation of the network and $p_c$ turns out to be $p_c \sim 0.64$.  Higher values of $\epsilon$, e.g. $\epsilon=0.4,0.6,0.8$ and $1.0$ provide us with much more dynamically robust network as then $p_c$ develops to $\sim 0.74,0.84,0.92$ and $1.0$ respectively. Therefore, increasing $\epsilon$ is quite capable of inducing global oscillation even when a large number of nodes have gone to the inactive mode. This result adds a significant perception to our existing knowledge of resuming dynamical robustness of damaged networks, more so when the networked system is exposed to long-range interactions among the dynamical units.

\begin{figure}
	\centerline{\includegraphics[scale=0.5400]{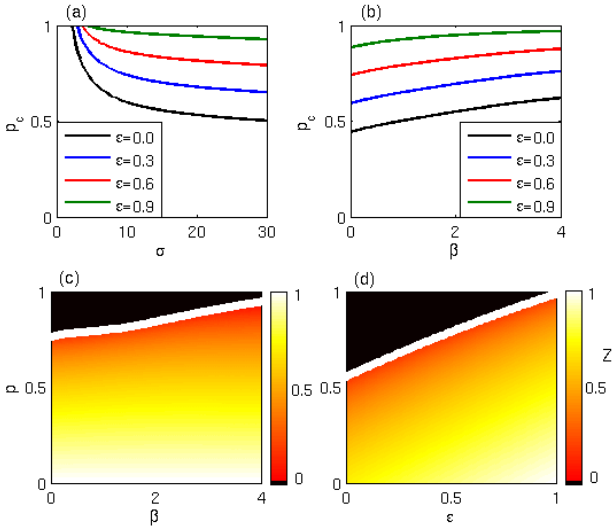}}
	\caption{The critical inactivation ratio $p_c$ for different values of the self-feedback strength $\epsilon$ with respect to : (a) interaction strength $\sigma$ (for $\beta=1$) and (b) decay rate $\beta$ (for $\sigma=30$). The order parameter $Z$ as a function of : (c) the inactivation ratio $p$ and the decay rate $\beta$ for which $\sigma=15$ is fixed and (d) the inactivation ratio $p$ and the self-feedback strength $\epsilon$ for which $\sigma=15$, $\beta=1$ are fixed.  Here $\langle k^{[1]} \rangle=20$ and $p_{r}=0.1$.  }
	\label{fig6}
\end{figure}

\par Looking at Fig. \ref{fig6}(a), one can understand better this scenario, particularly the variation in $p_c$ against increasing $\sigma$ (with $\beta=1$) for different $\epsilon$, whenever $\langle k^{[1]} \rangle=20$ and $p_{r}=0.1$. First, $p_c$ versus $\sigma$ is depicted whenever $\epsilon=0$, and the expected fall in the $p_c$ values is clear. Then we choose some non-zero $\epsilon$ and proceed. For example, $\epsilon=0.3$ results in a similar drop down but with higher values of $p_c$, implying enhancement of robustness. With higher $\epsilon=0.6$ and $0.9$, the $p_c$ curves shift upwards more (cf. Fig. \ref{fig6}(a)) indicating augmentation in robustness to greater extents. In a similar fashion, we plot  $p_c$ with respect to $\beta$ (with $\sigma=30$) in Fig. \ref{fig6}(b), where we first consider $\epsilon=0$, and then $\epsilon=0.3,0.6,0.9$. We have already perceived the fact that for increasing power-law exponent $\beta$, the robustness of the networked system also increases. Besides growing $p_c$ values for increasing $\beta$, it is considerably mentionable that $\epsilon$ keeps on uplifting the $p_c$ values for the whole range of $\beta$. These results suggest that as a consequence of the introduction of self-feedback, a notable improvement in the robustness against progressive dynamical inactivation of the nodes can be witnessed throughout any value of the parameters like $\sigma$ or $\beta$. Next, in Fig. \ref{fig6}(c) we plot $Z$ for varying $p$ and $\beta$ similar to Fig. \ref{fig4}(c), with the same other parameter values but this time for $\epsilon=0.5$ instead of $\epsilon=0$ (i.e., the no-feedback case). This introduction of $\epsilon$ helps in raising the values of $p_c$ quite effectively throughout the entire range of $\beta$. This results in a significant change in the phase diagram (compared to the one depicted in Fig. \ref{fig4}(c)) indicating notable enhancement in robustness of the system. Further, in Fig. \ref{fig6}(d), variation of $Z$ has been portrayed as a function of $p$ and the feedback strength $\epsilon$, whenever $\beta=1$. From this depiction, it can be easily witnessed how increasing $\epsilon$ leads to notable strictly monotonic raise in the $p_c$ values. In fact, for sufficient feedback ($\epsilon \sim 1$ here) the networked system becomes fully robust as $p_c$ turns out be $1$ for $\epsilon \sim 1$.   

\begin{figure}
	\centerline{\includegraphics[scale=0.4500]{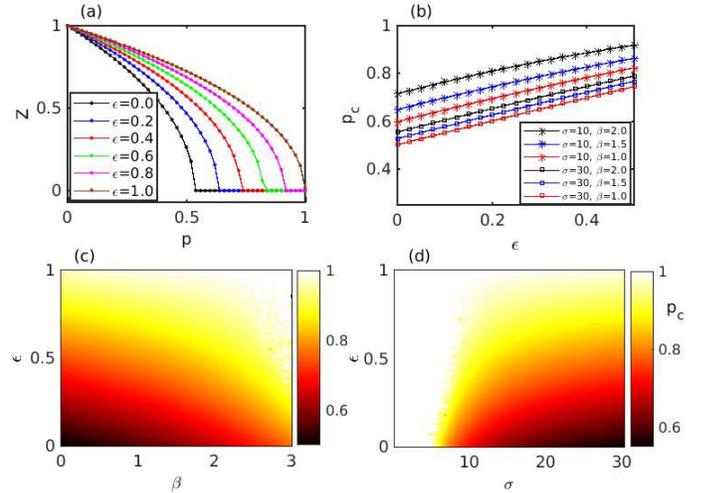}}
	\caption{(a) The order parameter $Z$ as a function of the inactivation ratio $p$ for different values of the self-feedback strength $\epsilon$ in case of the scale-free network model, for which $\beta=1$ and $\sigma=30$ are fixed; (b) The critical inactivation ratio $p_c$ with respect to $\epsilon$ for different sets of values of $\sigma$ and $\beta$. The critical $p_c$ depicted through the phase diagrams in the (c) $\epsilon-\beta$ (with $\sigma=15$ fixed), (d) $\epsilon-\sigma$ (with $\beta=2$ fixed) parameter planes. }
	\label{fig7}
\end{figure}

\par We then go for inspecting the impact of self-feedback imposed on dynamically damaged scale-free networks. In Fig. \ref{fig7}(a), we delineate the numerically simulated $Z$ against the inactivation ratio $p$ for fixed $\beta=1,~\sigma=30$ and different $\epsilon \in \{0,0.2,0.4,0.6,0.8,1.0\}$. As long as there is no feedback ($\epsilon=0$), $p_c \sim 0.54$. However, $\epsilon=0.2$ restrains the system to dynamically collapse that early, and increase $p_c$ to $p_c \sim 0.64$. Higher $\epsilon=0.4,0.6,0.8,1.0$ provide us with much more dynamically robust network, as then $p_c$ value advances to $\sim 0.74$, $\sim 0.84$, $\sim 0.92$ and $\sim 1.0$ respectively. Figure \ref{fig7}(b) discloses $p_c$ for definite values of $\beta$ and $\sigma$, depicted over a range of $\epsilon$. To start with, the $p_c$ values are plotted for $\sigma=10$ and $\beta=2.0,1.5,1.0$. Since self-feedback is basically making the network more resilient, $p_c$ values monotonically increase with increasing $\epsilon$, irrespective of the values of $\beta$. Then, we choose $\sigma=30$ with the same set of $\beta=2.0,1.5,1.0$ and as the results demonstrate, $\epsilon$ increases $p_c$ on the whole. However, the the networked system becomes more vulnerable to the dynamical damage compared to the previous situation as the interaction strength $\sigma$ is much higher this time. For a better understanding of the transition from aging to no-aging in the form of growing $p_c$ owing to the increasing $\epsilon$, the variation of $p_c$ in the $\beta-\epsilon$ parameter plane is shown in Fig. \ref{fig7}(c), where $\sigma=15$. The positive influence of increasing $\epsilon$ in resurgence of global oscillation for any $\beta$ is conspicuous from the figure. In addition, we outline $p_c$ for simultaneous change in $\sigma$ and $\epsilon$ (cf. Fig. \ref{fig7}(d)), in which $\beta=2$. Here also we observe the affirmative impact of $\epsilon$ in reviving dynamism throughout all $\sigma$. Thus, these outcomes clearly display the scenario of resumption of dynamic activity and hence the enhanced dynamical robustness of damaged complex networks comprising active and inactive units.      

\par Let us now mathematically contemplate with the situation in which all the nodes in the network (Eq. (\ref{eqq1})) are prone to self-feedback, then the dynamics on network is described as the following,
\begin{equation}
	\begin{array}{lcl}\label{eqq18}
		\dot{\bf x}_j={\bf f}({\bf x}_j)+\dfrac{1}{N}\sum\limits_{d=1}^{D}\sigma_d\sum\limits_{k=1}^{N}{A}^{[d]}_{jk}\Gamma({\bf x}_k-{\bf x}_j)+\epsilon{\bf x}_j,\\
		~~~~~~~~~~~~~~~~~~~~~~~~~~~~~~~~~~~j= 1,2,\cdots,N,
	\end{array}
\end{equation}
where $\epsilon$ accounts for the strength of the self-feedback, and all the other variables bear the same connotation as before. 
\par Then, with this new dynamical equations for SL oscillators (Eq. (\ref{eqq2})), the Eq. (\ref{eqq3}) transforms into 
\begin{equation}
	\begin{array}{lcl} \label{eqq19}
		\dot{z_j} = (\alpha + i\Omega - |z_j|^2)z_j + \dfrac{1}{N}\sum\limits_{d=1}^{D}\sigma_d{k_j}^{[d]}\Big[(1-p)M_A(t)\\~~~~~~~~~~~~~~~~~~~~~~~~~~~~~~~~~~~~~~+pM_I(t)-z_j\Big]+\epsilon z_j,\\
	\end{array}
\end{equation}
where the mean-fields (cf. Eqs.(\ref{eqq4}-\ref{eqq5})) remain the same, and further we obtain the amplitude evolution as  
\begin{equation}
	\small{\begin{array}{lcl}  \label{eqq20}
			\dot{r_j} = \Big(\alpha_j - \dfrac{1}{N}\sum\limits_{d=1}^{D}\sigma_d{k_j}^{[d]}-r_j^2+\epsilon\Big)r_j \\~~~~~~~~~~~~~~~~~~~~~~ +\dfrac{1}{N}\sum\limits_{d=1}^{D}\sigma_d{k_j}^{[d]}\Big[(1-p)R_A(t)+pR_I(t)\Big],\\
	\end{array}}
\end{equation}
in which $R_{A,I}(t)$ is given by Eq. (\ref{eqq7}). 
\par Now moving forward following the same procedure as above, it is easy to obtain the equation satisfied by the stationary amplitude $r_j^*$ as 
\begin{equation}
	\begin{array}{lcl} \label{eqq21}
		r_j^3 - \Big(\alpha_j-\dfrac{1}{N}\sum\limits_{d=1}^{D}\sigma_d{k_j}^{[d]}+\epsilon\Big)r_j\\
		~~~~~~~~~~~~~~~~~~ -\dfrac{1}{N}\sum\limits_{d=1}^{D}\sigma_d{k_j}^{[d]}\Big[((1-p)R_A+pR_I)\Big]=0,
	\end{array}
\end{equation}
where for unique positive real root one must have,  
\begin{equation}
	\begin{array}{lcl} \label{eqq22}
		\alpha_j - \dfrac{1}{N}\sum\limits_{d=1}^{D}\sigma_d{k_j}^{[d]}+\epsilon <0,~\forall  j \in S_A. 
	\end{array}
\end{equation}
One can then derive the new $(1,1)$-th entry of $J_0$ as 
\begin{equation}
	\begin{array}{lcl} \label{eqq23}
		\dfrac{\partial G_A}{\partial R_A}\Bigr\rvert_{R_A=R_I=0}\simeq
		\dfrac{\sum\limits_{d=1}^{D}\sigma_d\sum\limits_{j \in S_A}\Big[\dfrac{{k_j}^{[d]}L_j}{L_j-N\alpha_j-N\epsilon}\Big]}{N^2\sum\limits_{d=1}^{D}\sigma_d s^{[d]}}. 
	\end{array}
\end{equation}
Next, defining 
\begin{equation}
	\begin{array}{lcl} \label{eqq24}
		H^{\epsilon}(\sigma,\alpha)=\dfrac{\sum\limits_{d=1}^{D}\sigma_d\sum\limits_{j=1}^{N}\Big[\dfrac{{k_j}^{[d]}L_j}{L_j-N\alpha_j-N\epsilon}\Big]}{N^2\sum\limits_{d=1}^{D}\sigma_d s^{[d]}},
	\end{array}
\end{equation}
the new Jacobain matrix $J^{\epsilon}_0$ becomes
$$J^{\epsilon}_0=\begin{pmatrix}
	(1-p)H^{\epsilon}(\sigma,a) & pH^{\epsilon}(\sigma,a)\\
	~~(1-p)H^{\epsilon}(\sigma,-b) &~~ pH^{\epsilon}(\sigma,-b)
\end{pmatrix}.$$
Finally, the new critical inactivation ratio turns out to be
\begin{equation}
	\begin{array}{lcl} \label{eqq25}
		p^{\epsilon}_c = \dfrac{H^{\epsilon}(\sigma,a)-1}{H^{\epsilon}(\sigma,a)-H^{\epsilon}(\sigma,-b)}>p_c.
	\end{array}
\end{equation}
This expression of $p^{\epsilon}_c$ thus provides us with the critical inactivation ratio for the networked dynamical system subject to the scenario of self-feedback of strength $\epsilon$. We have used this formula while plotting the analytical curves over different phase diagrams above. \\

\section{Conclusion}
The modality of interactions in complex networks strongly determines how the networked systems function. It has significant influence in deciding the network's competence to withstand dynamical damages and also to recover afterwards. The issue of network robustness has emerged as one of the most crucial topic of recent research as there exists several natural instances in which complex systems are exposed to inevitable dynamical or structural disturbances. In this paper, we studied the dynamical robustness of complex damaged networks where the local units of the networks experience continuous dynamical degradation. Specifically, in the first part, we investigated what fraction of dynamically collapsed units leads to a exhaustive decline in the global oscillation of the whole network. We particularly emphasized here on networks subject to long-range interactions among the nodes. Through such format of connections, nodes interact not only via the direct short-range links, but also communicate on the basis of long-distant indirect pathways. For this, we examined diverse underlying network topologies from Erd\"{o}s-R\'{e}nyi random network to  Watts-Strogatz small world model and also Barab\'{a}si-Albert scale-free network. We have carried out thorough numerical computation and further validated all the outcomes with rigorous analytical result, that works no matter what network architecture is in the background of the system ensemble. In the second part of the manuscript, we confined ourselves in answering the question of protecting the network from the throughout collapse against such dynamical perturbations. In order to provide a simple yet highly potential mechanism that can effectively resurrect global rhythmicity in the network, we scrutinized the role of self-feedback of the local dynamical units induced into the system.  We proved both theoretically and numerically that, interestingly enough, only a self-feedback can efficiently resume dynamism in the network by a big margin.    Therefore, to increase the robustness of a networked system, adjusting feedback strength can be claimed to be good enough.





\section*{\normalsize{Acknowledgments}}
The author would like to thank Prof. Dibakar Ghosh for his valuable suggestions. The author also thanks the Planning and Budgeting Committee (PBC) of the Council for Higher Education, Israel for support.





\end{document}